\documentclass{nature}

\usepackage{caption}
\usepackage{graphicx}
\usepackage[scaled]{helvet}
\usepackage{amsmath}
\usepackage{amssymb}
\usepackage{bm}
\usepackage{xfrac}
\newcommand*{\myfont}{\fontfamily{phv}\selectfont}
\usepackage{hyperref}
\usepackage{anyfontsize}
\usepackage{ragged2e}

 \usepackage{color}

\definecolor{CLBlue}{rgb}{0, .25, .8}
\definecolor{MyBlue}{rgb}{0, .24, .40}
\definecolor{MyTurquoise}{rgb}{0, .53, .49}
\definecolor{MyGreen}{rgb}{0, .35, 0}
\definecolor{MyOrange}{rgb}{.8, .46, 0}
\definecolor{MyRed}{rgb}{.57, .07, 0}
\definecolor{MyPurple}{rgb}{.46, .1, .46}

\title{Irreversible behavior drives neural flows in the hippocampus}

\author{Kaiyue Shi$^{1,2,3,4}$ \& Christopher W. Lynn$^{3,4,5,*}$}

\begin{document}


\maketitle

\begin{affiliations}
\item Lewis-Sigler Institute for Integrative Genomics, Princeton University, Princeton, NJ 08544, USA
\item School of Physics, Peking University, Beijing 100871, China
\item Department of Physics, Yale University, New Haven, CT 06511, USA
\item Quantitative Biology Institute, Yale University, New Haven, CT 06511, USA
\item Wu Tsai Institute, Yale University, New Haven, CT 06510, USA
\item[*] Corresponding author: christopher.lynn@yale.edu
\end{affiliations}


\noindent {\large \myfont \textbf{Abstract}}
\vspace{-28pt}

\noindent\rule{\textwidth}{.5pt}

\noindent In the brain, neural activity undergoes directed flows between states, thus breaking time-reversal symmetry. At the same time, animals also exhibit irreversible flows between behavioral states. Yet it remains unclear whether---and how---irreversibility in the brain relates to irreversibility in behavior. Here, we explore this connection in the hippocampus, where neural activity encodes physical location. We show that hippocampal irreversibility can be quantified using the time-delayed cross-correlations between neurons. As a mouse moves along a virtual track, we find that physical flows through the animal's environment generate neural flows through its cognitive map. Strikingly, this neural irreversibility is explained by a minimal model with only three parameters: the average velocity of the mouse, the variance in this velocity, and the resolution of the neural encoding. Together, these results provide a mechanistic understanding of irreversibility in the hippocampus and shed light on the links between symmetry breaking in the brain and behavior.


\newpage

\noindent {\large \myfont \textbf{Main}}
\vspace{-28pt}

\noindent\rule{\textwidth}{.5pt}


Animals perform sequences of actions that are directed in time, from flocking and foraging to movement, communication, and social interactions.\cite{Gnesotto-01, Wang-02, Ferretti-01, Cavagna-01, Cavagna-02, Stephens-02, Stephens-01, Shannon-01, Lynn-07, Krause-01} Reverse the order of these actions, and the nature of each behavior changes entirely. In this way, animals break time-reversal symmetry by generating net flows between behavioral states.\cite{Gnesotto-01, Schrodinger-01} Similarly, recent evidence has established that neural dynamics also break time-reversal symmetry across scales. At the level of populations of neurons, collective dynamics exhibit irreversible flows between neural states.\cite{Lynn-11, Lynn-12} At the level of coarse-grained neural activity, irreversibility in the human brain is linked to cognition,\cite{Lynn-09, Deco-01} consciousness,\cite{Perl-01, delaFuente-01} and disease.\cite{Cruzat-01, Zanin-01} Yet despite this progress, we still lack a clear connection between irreversibility in the brain and behavior.

The hippocampus provides an ideal setting to investigate the breaking of time-reversal symmetry.\cite{OKeefe-01, OKeefe-03} A significant fraction of neurons in the hippocampus, known as place cells, each fire when an animal enters a specific location.\cite{OKeefe-02, Moser-01} Thus, the collective activity of place cells is thought to encode a cognitive map of the animal's environment, which is critical for spatial processing, episodic memory, and hippocampal replay.\cite{Morris-01, Bures-01, Smith-01, Foster-01, Buzsaki-01} Given this literal mapping from physical location to neural activity, a clear question emerges: As an animal moves through its environment, do irreversible flows in physical space give rise to irreversible flows in the cognitive map? \\

\noindent {\myfont \large Quantifying neural flows in the hippocampus}

To answer this question, we must first quantify the strength of neural flows in the hippocampus. When a place cell fires, this reflects a specific location within the cognitive map, thus defining a neural state.\cite{OKeefe-01, OKeefe-03, OKeefe-02, Moser-01} The rate of flow from one state $i$ to another state $j$ on a timescale $\tau$ is then given by $\frac{1}{\tau_\text{max} - \tau} \sum_{t = 1}^{\tau_\text{max} - \tau} x_i(t)x_j(t+\tau)$, where $x_i(t)$ represents the activity [$x_i(t) = 1$] or silence [$x_i(t) = 0$] of neuron $i$ at time $t$, and $\tau_\text{max}$ is the length of the recording. Notably, this flow is equivalent to the time-delayed cross-correlation between the two neurons,\cite{Aertsen-01}
\begin{equation}
\label{eq:Cij}
C_{ij}(\tau) = \langle x_i(t) x_j(t + \tau)\rangle_t,
\end{equation}
where $\langle \cdot \rangle_t$ represents an experimental average over time. If $C_{ij}(\tau) = C_{ji}(\tau)$ for all neurons $i$ and $j$---that is, if the correlations are symmetric---then there is no net flow between neural states; reverse the direction of time, and the dynamics remain identical. By contrast, any asymmetry in the correlations defines a net flow $C_{ij}(\tau) - C_{ji}(\tau) \neq 0$ between two locations in the cognitive map, thus breaking time-reversal symmetry.





\begin{figure}[t!]
\centering
\includegraphics[width= \textwidth]{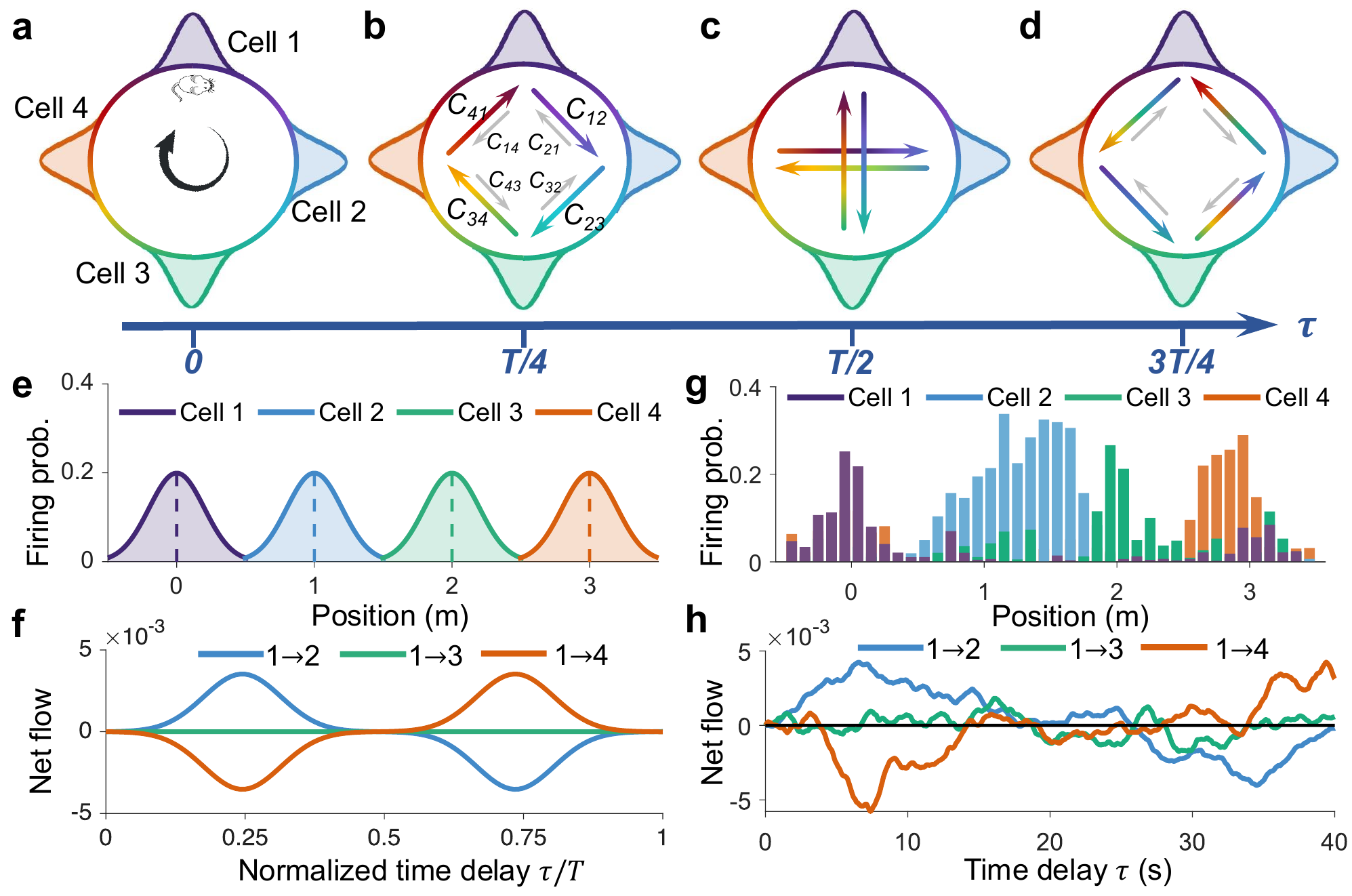}
\raggedright
\captionsetup{labelformat=empty}
{\spacing{1.25} \caption{\small \textbf{Fig.~1 $|$ Flows between neural states in the hippocampus.}
\textbf{a}, Place fields for four hypothetical place cells spaced evenly around a circular track as a mouse runs with constant velocity and period $T$. \textbf{b-d}, Flows between neural states $C_{ij}(\tau)$ for different timescales $\tau$, with grey arrows indicating negligible flows. As the mouse advances along the track, flows progress from clockwise (\textbf{b}) to directly across the track (\textbf{c}) to counterclockwise (\textbf{d}).
\textbf{e}, Firing probabilities for the hypothetical cells in \textbf{a}-\textbf{d}, defined by Gaussians (with uniform variance) spaced evenly along a four-meter track. \textbf{f}, Net flows $C_{ij}(\tau) - C_{ji}(\tau)$ from cell 1 to cells 2, 3, and 4 as functions of the normalized time delay $\tau/T$.
\textbf{g}, Place fields for four neurons recorded experimentally from the hippocampus of a mouse as it runs along virtual track of length $L = 4$~m.
\textbf{h}, Net flows $C_{ij}(\tau) - C_{ji}(\tau)$ from cell 1 to cells 2, 3, and 4 in \textbf{g} as functions of the time delay $\tau$.
\label{fig:1}}}
\end{figure}

To illustrate these neural flows, consider a mouse running around a circular track. The irreversibility of the mouse's behavior is self-evident: reverse the sequence of positions, and the mouse runs in the opposite direction. Now consider four place cells that fire at locations spaced evenly along the track, with firing rates represented by the place fields in Fig.~\ref{fig:1}a. If the mouse runs with period $T$, then it moves from one place field to the next on a timescale $\tau = T/4$. This physical movement induces net flows between neural states, breaking time-reversal symmetry in the cognitive map (Fig.~\ref{fig:1}b). As the timescale increases to a half-period $\tau = T/2$, the flows between different neurons cancel, yielding dynamics that appear entirely reversible (Fig.~\ref{fig:1}c). Increase the timescale further to $\tau = 3T/4$, and net flows emerge once again, but in the direction opposite the mouse's movement (Fig.~\ref{fig:1}d). We therefore arrive at the intuition that (i) irreversible movement through an environment may drive irreversible flows in the hippocampus, and (ii) these flows should depend critically on the timescale $\tau$ (Fig.~\ref{fig:1}f). \\

\noindent {\myfont \large Irreversibility in the hippocampus}

To test these intuitions, we investigate the activity of $1485$ neurons in the CA1 region of the mouse hippocampus, recorded using two-photon calcium imaging (Methods).\cite{Gauthier-01, Meshulam-01} The mouse runs with its head fixed along a virtual circular track of length $L = 4\,\text{m}$. Among all neurons, we identify $462$ ($31\%$) with statistically significant place fields (Methods), consistent with previous estimates for the density of place cells.\cite{OKeefe-01, Meshulam-01} Consider four such cells with place fields centered approximately equal distances along the track (Fig.~\ref{fig:1}g), thus mirroring our previous example (Fig.~\ref{fig:1}e). From one cell to the next (in the direction of the mouse's trajectory), we observe an initial peak in the net flow $C_{ij}(\tau) - C_{ji}(\tau)$ followed by a gradual decline to negative flow as $\tau$ increases (Fig.~\ref{fig:1}h, blue). Precisely as expected (Fig.~\ref{fig:1}f), this pattern is directly inverted for cells in the reverse direction (Fig.~\ref{fig:1}h, orange), and we observe no net flows between cells located on opposite sides of the track (Fig.~\ref{fig:1}h, green).



Rather than studying the flow between each pair of neurons individually, we can quantify the total strength of flows combined across all neurons in the population,
\begin{equation}
    \label{eq:Sigma}
    \Sigma(\tau) = \frac{1}{2} \sum_{i,j} (C_{ij}(\tau) - C_{ji}(\tau))^2,
\end{equation}
where the factor of $1/2$ avoids double counting. If the neural dynamics obey time-reversal symmetry on timescale $\tau$, then the net flows between all neurons vanish, and $\Sigma(\tau) = 0$. Meanwhile, any net flow $C_{ij}(\tau) - C_{ji}(\tau) \neq 0$ between a pair of neurons leads to an increase in the total flow $\Sigma(\tau)$. In fact, we show that $\Sigma$ serves as a lower bound on the information-theoretic irreversibility, which plays a fundamental role in non-equilibrium statistical physics (Methods).\cite{Seifert-01, Roldan-01, Cover-01, Lynn-11} Finally, note that with no time delay ($\tau = 0$), one will never observe flows in any system, such that $\Sigma (\tau = 0) = 0$ by definition.

\begin{figure}[t]
\centering
\includegraphics[width=.9\textwidth]{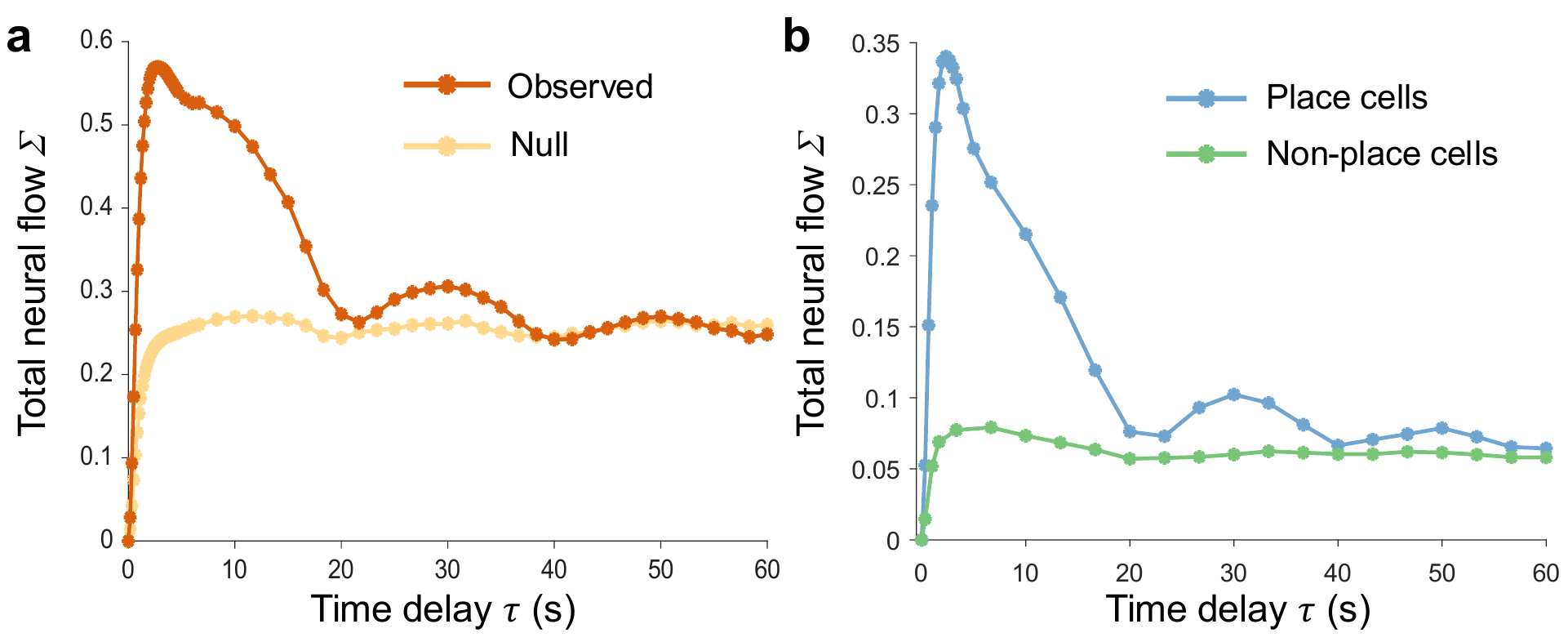}
\captionsetup{labelformat=empty}
{\spacing{1.25} \caption{\small \textbf{Fig.~2 $|$ Neural flow in the hippocampus across timescales.}
\textbf{a}, Total neural flow $\Sigma(\tau)$ across a population of $1485$ neurons in the hippocampus of a mouse running along a circular track (Methods).\cite{Gauthier-01, Meshulam-01} Null values are computed for the same population, but with the activity time-series of each neuron shifted by a random length of time (Methods). \textbf{b}, Total neural flow $\Sigma(\tau)$ among the subpopulation of $462$ place cells and among the remaining $1023$ non-place cells.
\label{fig:2}
}}
\end{figure}

Quantifying the total flow between all neurons, we discover that neural dynamics in the hippocampus are highly irreversible (Fig.~\ref{fig:2}a). Compared to null neurons with activity randomly shifted in time, the real neurons exhibit flows that are up to twice as strong. Moreover, this irreversibility displays clear patterns on three distinct timescales. On a short timescale, the total flow undergoes a sharp increase, reaching a maximum at $\tau = 2.7\,\text{s}$. On an intermediate timescale, the neural flows appear to oscillate with a period of $\tau \approx 20\,\text{s}$. Finally, on a longer timescale of $\tau \approx 1\,\text{min}$, the overall strength of flows decays to the baseline value of the null population. Focusing specifically on the place cells, we find that these patterns are exaggerated (Fig.~\ref{fig:2}a): the peak in the neural flow becomes sharper, the oscillations are clearer, and the decay is stronger. By contrast, the flows between non-place cells are significantly weaker and display none of these characteristic features (Fig.~\ref{fig:2}b). Together, these results demonstrate that irreversibility in the hippocampus arises primarily from the dynamics of place cells. But how are these neural flows driven by behavior? \\


\noindent {\myfont \large Oscillations in neural flow driven by velocity}

\begin{figure}[t]
\centering
\includegraphics[width= .95\textwidth]{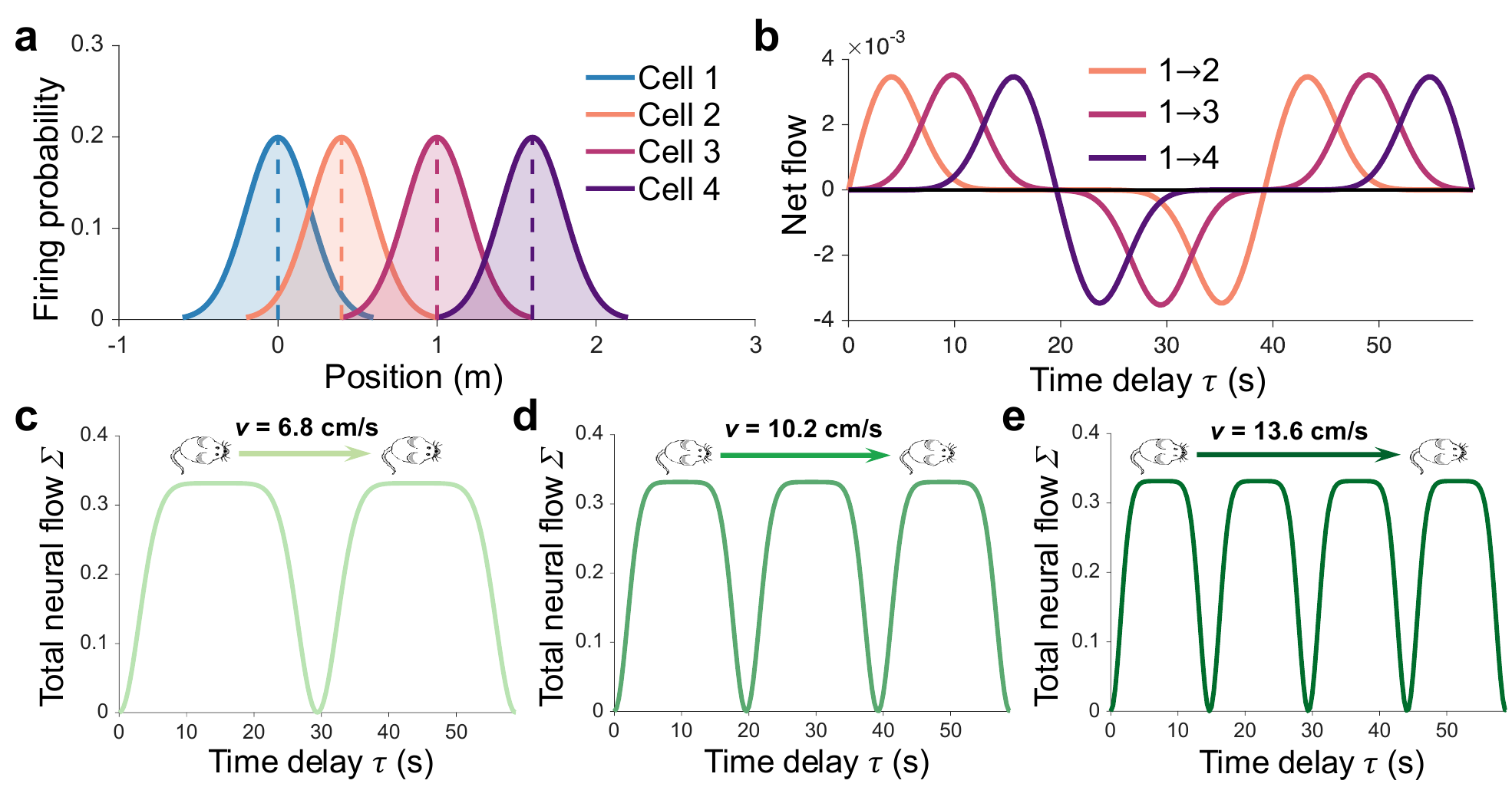}
\captionsetup{labelformat=empty}
{\spacing{1.25} \caption{\small \textbf{Fig.~3 $|$ Mouse velocity determines period of oscillations in neural flow.}
\textbf{a}, Example of four place cells in our model, each with a place field defined by a Gaussian centered at a random location along a track of length $L = 4\,\text{m}$ with a common standard deviation $\sigma = 20\,\text{cm}$. \textbf{b}, Net flows $C_{ij}(\tau) - C_{ji}(\tau)$ from cell 1 to the other three cells for a mouse moving at a constant velocity $v = 10\,\text{cm/s}$, yielding a period $T = 40\,\text{s}$. The net flows between all cells vanish for multiples of $\tau=\frac{T}{2}$, the half-period of the mouse's movement. \textbf{c-e}, Total neural flow $\Sigma(\tau)$ between $462$ synthetic place cells for mice running at different velocities $v$. For $v=10.2\,\text{cm/s}$ (\textbf{d}), the period of oscillations matches that measured in experiments (Fig.~\ref{fig:2}).
\label{fig:3}
}}
\end{figure}

To understand how an animal's physical movement drives neural flows, we construct a minimal model connecting the two. Consider a population of place cells, each with a place field that defines the firing probability as a function of location. For simplicity, the place field for each cell $i$ is defined by a Gaussian centered at a random position $\mu_i$ along the circular track (Fig.~\ref{fig:3}a). The standard deviation $\sigma$, which we hold constant across cells, defines the resolution with which place cells encode the animal's location. If the mouse runs with constant velocity $v$, then it completes one lap around the track with period $T=\frac{L}{v}$, where $L$ is the track length. For a given pair of cells $i$ and $j$, we expect the net flow $C_{ij}(\tau) - C_{ji}(\tau)$ to be maximized at the time it takes the mouse to travel from $\mu_i$ to $\mu_j$ (that is, $\tau = \frac{\mu_j - \mu_i}{v}$ if $\mu_j \ge \mu_i$); conversely, the net flow should be minimized for $\tau = T - \frac{\mu_j - \mu_i}{v}$. Indeed, calculating the net flows (Methods), we observe precisely these maxima and minima, with the pattern repeating for each period $T$ (Fig.~\ref{fig:3}b).

Across all pairs of cells, we find that the net flows vanish at multiples of the half-period $\tau = \frac{T}{2}, T, \frac{3T}{2},\hdots$ (Fig.~\ref{fig:3}b). This reversibility in the neural dynamics is directly related to the reversibility of movement in physical space. If we take a snapshot of the mouse's location every half-period, then it appears to complete half a lap at each step, and the trajectory is identical whether viewed forward or backward in time. Given this reversibility in behavior, our model predicts that the total flow $\Sigma(\tau)$ combined across an entire group of place cells should oscillate with a period half that of the animal's physical movement (Fig.~\ref{fig:3}c-e). Thus, as the mouse runs faster, the total neural flow oscillates with increasing frequency. For a track of length $L = 4\,\text{m}$ and oscillations with period $\tau \approx 20\,\text{s}$ (as observed in Fig.~\ref{fig:2}), our model predicts a velocity of $v = \frac{L}{2T} \approx 10\,\text{cm/s}$ that matches the average speed of the mouse $v = 10.2\,\text{cm/s}$ in the experiment  (Fig.~\ref{fig:3}d; Methods). \\

\noindent {\myfont \large Decay in neural flow driven by diffusion}

We have found that the oscillations in neural flow are driven by periodicity in behavior, but our model does not yet account for the decay in irreversibility nor the sharp peak (Fig.~\ref{fig:2}). One might expect the decay in neural flows to arise from noise or heterogeneity in the place cells. However, simulations reveal that this decay cannot be explained by noise in the place fields, differences in firing rates, heterogeneity in place field widths, or correlations in place field locations (Supplementary Information). Ruling out these potential neural mechanisms, we turn to variability in behavior. Rather than running at a constant velocity, the speed of the mouse naturally fluctuates, which leads to its physical location becoming de-correlated over time. If this physical de-correlation drives de-correlation in neural activity, then we should expect the neural flows to decay on long timescales.

\begin{figure}[t!]
\centering
\includegraphics[width= \textwidth]{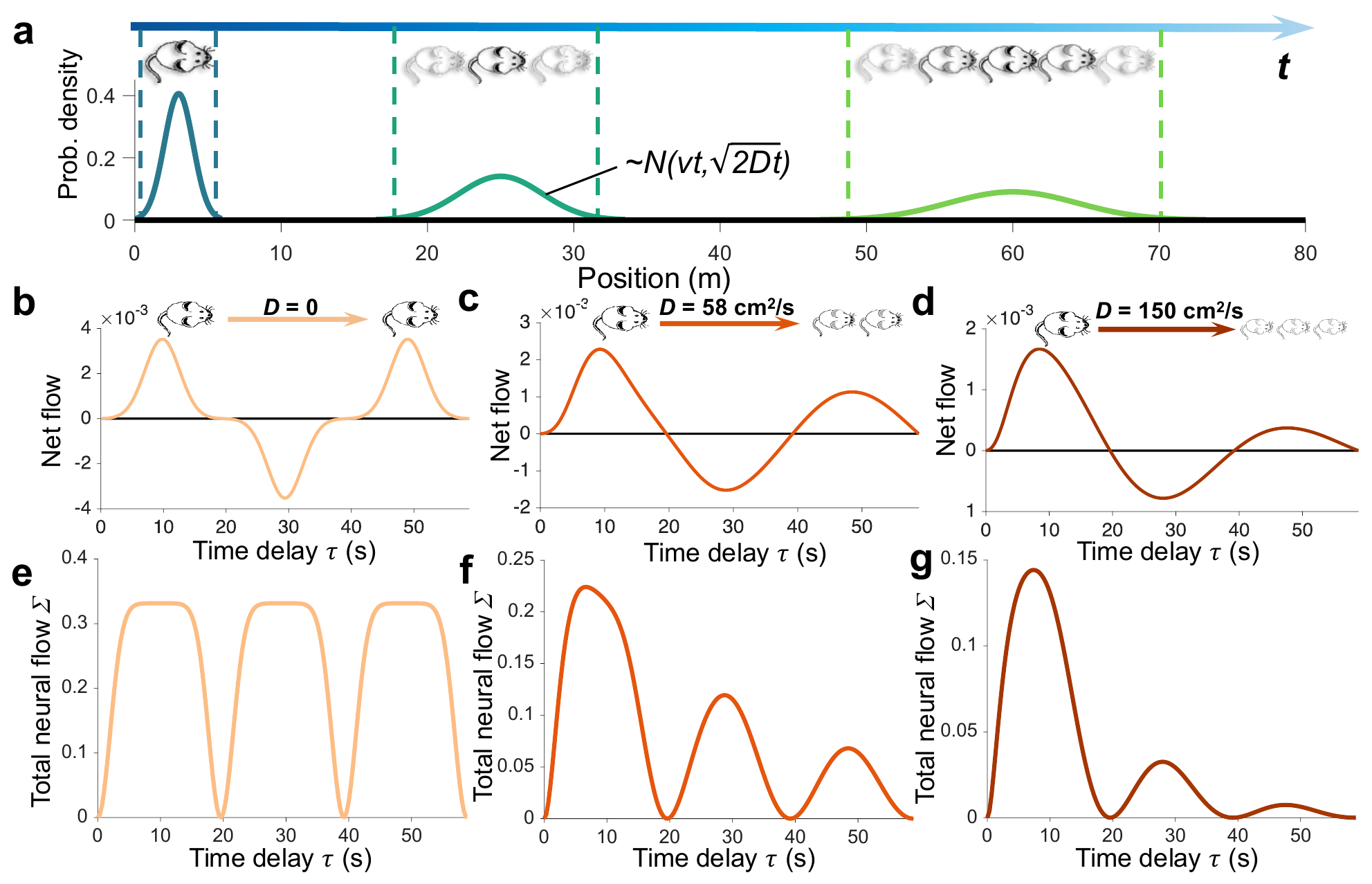}
\captionsetup{labelformat=empty}
{\spacing{1.25} \caption{\small \textbf{Fig.~4 $|$ Diffusion in movement leads to decay in neural flow.}
\textbf{a}, Illustration of a biased random walk in which the distance traveled in time $t$ is given by a Gaussian $\mathcal{N}(vt, 2Dt)$, where $v$ is the mean velocity of the mouse and $D$ is the diffusion coefficient.
\textbf{b-d}, Net flow between two neurons with place fields centered a distance $\frac{L}{4}$ apart for different diffusion coefficients $D$. \textbf{e-g}, As $D$ increases, the total flow across an entire population $\Sigma(\tau)$ decays more rapidly. Experimental behavior is best described by $D=58\,\text{cm}^2$/s (\textbf{c} and \textbf{f}). Across all panels, calculations are performed by numerically solving the model with place field width $\sigma = 20\,\text{cm}$ and average velocity $v=10.2\,\text{cm/s}$ (Methods).
\label{fig:4}
}}
\end{figure}

To test this hypothesis, we extend our model to include variation in mouse velocity. Specifically, we model the movement of the mouse as a biased random walk with mean velocity $v$ and diffusion coefficient $D$.\cite{Codling-01} Thus, the distance that the mouse travels after an amount of time $t$ is described by a Gaussian $\mathcal{N}(vt, 2Dt)$ that grows wider as $t$ increases (Fig.~\ref{fig:4}a). With no diffusion ($D = 0$), the mouse runs at a constant velocity, as in our original model (Fig.~\ref{fig:3}). In this limit, the place cells remain correlated over time (Fig.~\ref{fig:4}b), and we observe no decay in neural flows (Fig.~\ref{fig:4}e). As diffusion increases to the value $D = 58\,\text{cm}^2$/s that best describes the experimental behavior of the mouse (Methods), we find that the place cells become de-correlated over time (Fig.~\ref{fig:4}c). Combined across an entire population, this leads the total strength of flows $\Sigma(\tau)$ to decrease with increasing time-delay $\tau$. As the animal's behavior becomes even more random (increasing $D$), this decay in neural irreversibility grows stronger (Fig.~\ref{fig:4}d,g). Note that despite this decay, the total flow between neurons still oscillates with a period $T/2$, independent from the strength of diffusion $D$. We therefore find that both the oscillations and decay in neural irreversibility are determined by specific features of the mouse's behavior: average velocity and diffusion. \\

\noindent {\myfont \large Maximum neural flow driven by resolution}

\begin{figure}[t!]
\centering
\includegraphics[width= 0.8\textwidth]{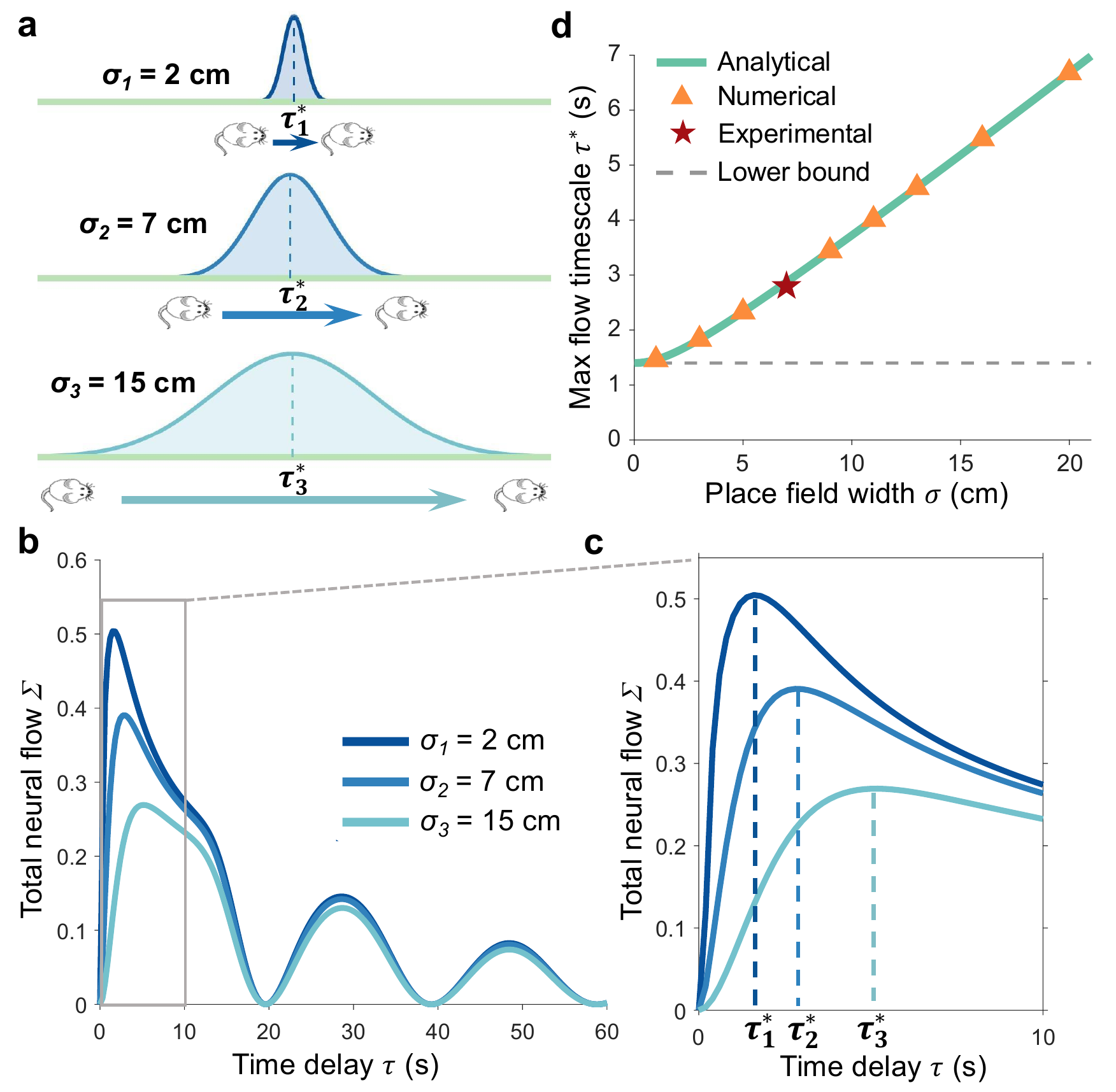}
\captionsetup{labelformat=empty}
{\spacing{1.25} \caption{\small \textbf{Fig.~5 $|$ Resolution of place cells determines peak in neural flow.}
\textbf{a}, For wider place fields (larger $\sigma$), animals with the same behavior will take longer to travel from one place field to the next, which is required to generate neural flows. \textbf{b-c}, Total neural flow $\Sigma(\tau)$ as a function of time-delay $\tau$ for different place field widths $\sigma$. As $\sigma$ decreases, a peak appears with a higher maximum flow on a shorter timescale $\tau^*$. \textbf{d}, Timescale of maximum neural flow $\tau^*$ as a function of place field width $\sigma$. Solid line is the analytic prediction in Eq.~(\ref{eq:tau_sigma}) (Methods), orange points are computed numerically, and dashed line represents the minimum value $\tau^*_\text{min}$ for perfect resolution ($\sigma = 0$). Red point indicates the timescale $\tau^* = 2.7\,\text{s}$ and average place field width $\sigma \approx 7\,\text{cm}$ measured in experiment (Methods). Across all panels, the velocity and diffusion of the mouse's movement are set to the experimental values $v = 10.2\,\text{cm/s}$ and $D = 58\,\text{cm}^2$/s.
\label{fig:5}
}}
\end{figure}

Thus far, our explanations for neural flows have not relied on the properties of the place cells themselves. On short timescales, however, we observe a sharp peak in the irreversibility that is not yet captured by our model (Fig.~\ref{fig:2}). To understand the timescale $\tau^*$ associated with this peak, we must consider the width of the place fields $\sigma$, which defines the resolution of the cognitive map. For large $\sigma$, it takes longer for the animal to travel from one place field to the next, which is required to produce net flows between neurons (Fig.~\ref{fig:5}a). By contrast, for small $\sigma$ the animal will spend less time crossing each place field, and neural flows should increase sharply for short time-delays $\tau$.


Decreasing the place field width $\sigma$, we find that a pronounced peak emerges in the total neural flow at small $\tau$ (Fig.~\ref{fig:5}b). As the place fields become narrower, increasing their spatial resolution, the maximum flow increases and the associated timescale $\tau^*$ decreases (Fig.~\ref{fig:5}c). Thus, using a minimal model with only three parameters---velocity $v$, diffusion $D$, and place field width $\sigma$---one can explain all of the key features of the neural flows measured in the hippocampus (Fig.~\ref{fig:2}). Moreover, we confirm that the same neural flows emerge in simulations of mouse behavior and neural activity (Supplementary Information).

For short time-delays $\tau$, the model can be solved analytically, yielding an equation relating the timescale of maximum flow $\tau^*$ to the three parameters $v$, $D$, and $\sigma$ (Methods). Holding the behavior of the mouse fixed (with $v$ and $D$ set to their experimental values) and sweeping over $\sigma$, we confirm that the analytic prediction for $\tau^*$ matches numerical calculations (Fig.~\ref{fig:5}d). In the limit of perfect neural resolution ($\sigma = 0$), we find that the peak timescale $\tau^*$ does not approach zero; instead, it settles to a minimum value $\tau^*_\text{min} = \lambda \frac{D}{v^2}$, where $\lambda = 2.51$ is a constant (Fig.~\ref{fig:5}d). This minimum $\tau^*_\text{min}$ defines the timescale at which the irreversibility of the behavior itself is maximized, without any reference to the cognitive map. Finally, using the analytic relationship between $\tau^*$ and $\sigma$ in our model, we can solve for the place field width that produces the peak timescale $\tau^* = 2.7\,\text{s}$ observed in experiment (Fig.~\ref{fig:2}). Remarkably, the model predictions are consistent with place fields of width $\sigma \approx 7\,\text{cm}$, which matches the average width measured in our population of place cells (Fig.~\ref{fig:5}d). \\

\noindent {\myfont \large Discussion}

Despite growing evidence that the brain breaks time-reversal symmetry across scales, animals, and neural systems,\cite{Lynn-11, Lynn-12, Lynn-09, Deco-01, Perl-01, delaFuente-01, Cruzat-01, Zanin-01} a concrete connection to behavior has yet to be established. We investigate this connection in the hippocampus, where neural activity defines a cognitive map of an animal's environment.\cite{OKeefe-01, OKeefe-03, OKeefe-02, Moser-01} Quantifying the flows between neural states, we discover that the hippocampus is highly irreversible. A minimal model directly links this neural irreversibility to animal behavior: as a mouse runs along a circular track, the movement of the animal through physical space generates irreversible flows between neural states in the cognitive map. Using only three parameters---mouse velocity, mouse diffusion, and place cell resolution---this model explains the neural irreversibility measured experimentally in the hippocampus.

Given the simplicity of our model, future work can immediately introduce additional biological realism. For example, while we treat place fields as distributed randomly with uniform widths, the firing of place cells can be highly heterogeneous.\cite{Hayman-01, Anderson-01} We note, however, that place cell heterogeneity does not explain the observed patterns in the neural flows (Supplementary Information). Similarly, while we model mouse movement as a biased random walk, animal behavior is known to exhibit complex correlations across timescales.\cite{Bialek-03, Stephens-02, Leighton-01} The fact that these details are not needed to understand the observed neural irreversibility demonstrates the surprising simplicity of our results. However, investigations into the fine-grained flows between specific neurons will likely require additional biophysical complexity.

More generally, we have only scratched the surface of understanding how external variables drive irreversibility in the brain. As animals move through more complicated environments, are these flows still mirrored in the hippocampus? One can ask similar questions about sequences of abstract concepts and memories, which are also thought to be encoded in the hippocampus.\cite{Buzsaki-01, Fortin-01} In other neural systems, collective activity is tightly linked to motor movements, sensory stimuli, and social processing.\cite{Svoboda-01, Stringer-01, Leonard-01, Fulton-01, Chen-03} Does irreversibility in these domains also drive neural irreversibility? Or, alternatively, do neural flows in some systems emerge endogenously from interactions between neurons? Our work provides hope for simple answers to these questions.

\newpage

\noindent {\large \myfont \textbf{Methods}}
\vspace{-28pt}

\noindent\rule{\textwidth}{.5pt}

\begin{methods}

\subsection{Neural data.} We study patterns of activity in $1485$ neurons in the hippocampus of a mouse, recorded in a recent experiment.\cite{Gauthier-01} The mouse is genetically modified so that its neurons express a protein whose fluorescence is modulated by calcium concentration, which in turn follows the electrical activity of the cells. This fluorescence is recorded using a scanning two--photon microscope as the mouse runs along a virtual track of length $L = 4\,\text{m}$. The signal from each cell $i$ consists of a quiet background punctuated by short bursts of activity,\cite{Meshulam-01} providing a natural binarization into active [$x_i(t) = 1$] or silent [$x_i(t) = 0$] within each video frame $t$. Capturing images at $30\,\text{Hz}$ for $39\,\text{min}$ yields $\tau_\text{max} = 7.3\times 10^4$ samples of activity. To generate null data with neural activity that is decoupled from animal position, we shift the time-series for each neuron by a random length of time (Fig.~\ref{fig:2}a). For each neuron, this is equivalent to a random circular permutation in time.

\subsection{Identifying place cells.} To identify place cells within the population, we first divide the track into 40 bins of equal width and calculate the probability of activity for each neuron conditional on the animal being located within each bin. For each neuron, we then generate null time-series as described above, randomly rotating the activity in time.\cite{Meshulam-01} We repeat this process 1000 times for each neuron and compute the conditional firing probabilities across bins for each permutation. We identify a neuron as a place cell if there are at least three consecutive bins in which the true conditional firing probability is above the $95^\text{th}$ percentile across random permutations. This procedure identifies 462 place cells out of 1485 neurons in total ($31\%$ of the population), consistent with previous estimates for the density of place cells.\cite{Meshulam-01, OKeefe-01}

\subsection{Total flow and irreversibility.} Here, we show that the total neural flow in Eq.~(\ref{eq:Sigma}) is a lower bound on the information-theoretic irreversibility, which plays a fundamental role in non-equilibrium statistical physics.\cite{Seifert-01, Roldan-01, Cover-01, Lynn-11} We define the state of the system based on the most recent neuron to fire, such that the number of states matches the number of neurons. The flow from state $i$ to state $j$ on timescale $\tau$ is given by the time-delayed cross-correlation $C_{ij}(\tau)$ in Eq.~(\ref{eq:Cij}). Given a set of flows $C_{ij}(\tau)$, the information-theoretic irreversibility is the Kullback-Leibler (KL) divergence between the forward- and reverse-time flows\cite{Roldan-01, Cover-01, Lynn-11, Lynn-12, Lynn-09}
\begin{equation}
    \label{eq:I}
    I(\tau) = \frac{1}{2} \sum_{i,j}(C_{ij}(\tau)-C_{ji}(\tau)) \ln\frac{C_{ij}(\tau)}{C_{ji}(\tau)},
\end{equation}
where the factor $1/2$ avoids double counting. This irreversibility measures the extent to which the dynamics break detailed balance on timescale $\tau$, such that $C_{ij}(\tau) \neq C_{ji}(\tau)$. Note that the flows $C_{ij}(\tau)$ are often normalized to define transition rates or joint transition probabilities; these choices of normalization simply re-scale $I(\tau)$ by a constant.\cite{Lynn-11, Lynn-12, Yu-01} In what follows, we normalize the flows such that $\sum_{ij} C_{ij}(\tau) = 1$; that is, we normalize the flows to define joint transition probabilities.

For a given pair of neurons $i$ and $j$,  let $a= \max \left\{ C_{ij},C_{ji} \right\}$ and $b= \min \left\{ C_{ij},C_{ji} \right\}$, such that $0< b \leq a < 1$. The corresponding term in Eq.~(\ref{eq:I}) becomes $(a-b)\ln \frac{a}{b}$. Using the fact that $\ln (1+x) \geq \frac{x}{x+1}$ for any $x$, we have $\ln \frac{a}{b}=\ln (1+\frac{a-b}{b}) \geq \frac{a-b}{a}$. Thus, each term in $I(\tau)$ can be lower-bounded, $(a-b)\ln \frac{a}{b} \geq \frac{(a-b)^2}{a}>(a-b)^2$. Combining across all pairs of neurons, we have,
\begin{equation}
    \label{eq:lower_bound}
    I(\tau) = \frac{1}{2} \sum_{i,j}(C_{ij}(\tau)-C_{ji}(\tau)) \ln\frac{C_{ij}(\tau)}{C_{ji}(\tau)} > \frac{1}{2} \sum_{i,j} (C_{ij}(\tau)-C_{ji}(\tau))^2 = \Sigma(\tau).
\end{equation}
We therefore find that the total flow $\Sigma(\tau)$ in Eq.~(\ref{eq:Sigma}) provides a lower bound on the information-theoretic irreversibility $I(\tau)$.




\subsection{Analytical model solution.}

We model a population of place cells, where the conditional firing rate (that is, the place field) of each neuron $i$ is proportional to a Gaussian $\mathcal{N}(\mu_i, \sigma^2)$. Explicitly, the firing rate of cell $i$ at position $s$ along the track is given by
\begin{equation}
r_i(s) = \frac{1}{k\sqrt{2\pi}\sigma}e^{-\frac{(s-\mu_i)^2}{2\sigma^2}},
\end{equation}
where $\mu_i$ is the place field center, $\sigma$ is the place field width (held constant across all cells), and $k$ is a constant that determines the overall firing rate. Note that the firing rate is defined periodically for positions $\mu_i - \frac{L}{2} + nL \le s \le \mu_i + \frac{L}{2} + nL$, where $n$ is any integer. Adjusting $k$ simply re-scales all of our results by a constant factor; we use $k = 10$ for all calculations. We model the movement of the mouse as a biased random walk with average velocity $v$ and diffusion coefficient $D$. Thus, the distance $d$ that the mouse travels after time-delay $\tau$ is also Gaussian distributed,
\begin{equation}
P_\tau(d) = \frac{1}{\sqrt{4\pi D\tau}} e^{-\frac{(d-v\tau)^2}{4D\tau}}.
\end{equation}

To compute the time-delayed cross-correlation $C_{ij}(\tau)$, we assume (without loss of generality) that $\mu_i = 0$ and $\mu_j = \delta$ (where $0\le \delta \le L/2$). Integrating over starting positions $-\frac{L}{2} \le s_i \le \frac{L}{2}$ (at time zero) and all possible final positions $s_j$ (at time $\tau$), we have
\begin{align}
\label{eq:Cij_int}
C_{ij}(\tau) &= \int_{-\frac{L}{2}}^\frac{L}{2} ds_i \int_{\delta - \frac{L}{2}}^{\delta + \frac{L}{2}} ds_j \sum_{n = -\infty}^\infty \frac{1}{L}P_\tau(s_j - s_i + nL) r_i(s_i)r_j(s_j),
\end{align}
where the sum in the second equation accounts for the possibility that the mouse travels $n$ laps forward or backward along the track. The physical meaning of each term is clear: The first two terms are the joint probability that the animal is at position $s_i$ at time $t = 0$ and $s_j + nL$ at $t+\tau$; the third term is the firing rate of neuron $i$ at $s_i$; and the fourth term is the firing rate of neuron $j$ at $s_j$ (which is equivalent to the firing rate at $s_j + nL$).

Each term in the sum in Eq.~(\ref{eq:Cij_int}) defines the contribution to $C_{ij}(\tau)$ from the mouse traveling $n$ laps around the track. On short timescales $\tau$, the mouse is unlikely to travel one lap in either direction, so we can focus on $n = 0$. Additionally, because the place fields are narrow relative to the length of the track, we can approximate the integrals by taking their bounds to infinity, yielding
\begin{align}
C_{ij}(\tau) &\approx \int_{-\infty}^\infty ds_i \int_{-\infty}^\infty ds_j \frac{1}{L}P_\tau(s_j - s_i) r_i(s_i)r_j(s_j), \\
&= \frac{1}{Lk^2 \sqrt{2\pi}\sigma_\tau}e^{-\frac{(v\tau - \delta)^2}{2\sigma_\tau^2}},
\end{align}
where $\sigma_\tau = \sqrt{2\sigma^2 + 2D\tau}$. We therefore arrive at an analytic expression for the net flow at short times,
\begin{equation}
C_{ij}(\tau) - C_{ji}(\tau) \approx \frac{1}{k^2 L\sqrt{2\pi}\sigma_\tau} \left( e^{-\frac{(v\tau-\delta)^2}{2\sigma_\tau^2}} - e^{-\frac{(v\tau+\delta)^2}{2\sigma_\tau^2}}\right).
\end{equation}

To compute the total neural flow $\Sigma(\tau)$, we average the net flow over the possible differences in place field locations $\delta$ and multiply by the number of neuron pairs. Thus, on short timescales, we arrive at the total neural flow,
\begin{align}
    \Sigma(\tau) &\approx \frac{N(N-1)}{2} \int_0^{\frac{L}{2}} d\delta \frac{2}{L}\left( C_{ij}(\tau)-C_{ji}(\tau)\right)^2 \\
    \label{eq:total_flow}
    &=\frac{A}{\sqrt{\sigma^2+D\tau}} \left(1-e^{-\frac{v^2\tau^2}{2(\sigma^2+D\tau)}}\right),
\end{align}
where $A = \frac{N(N-1)}{2\sqrt{2\pi}k^4L^3}$ is a constant. The maximum total flow is defined by the condition $\frac{\partial \Sigma}{\partial \tau} = 0$, from which we obtain an analytical relationship between the maximum flow timescale $\tau^*$ and the parameters in the model,
\begin{equation}
    \label{eq:tau_sigma}
    \left(1+\frac{v^2\tau^*(2\sigma^2+D\tau^*)}{D(\sigma^2+D\tau^*)}\right)e^{-\frac{v^2{\tau^*}^2}{2(\sigma^2+D\tau^*)}}=1.
\end{equation}
For general $v$, $D$, and $\sigma$, we cannot solve for $\tau^*$ explicitly. However, in the limit of perfect place cell resolution ($\sigma = 0$), we have the minimum timescale
\begin{equation}
\tau^*_\text{min} = \lambda \frac{D}{v^2},
\end{equation}
where $\lambda = 2.51$ is a constant (Fig.~\ref{fig:5}d).

\subsection{Numerical model solution.}

To compute the correlations $C_{ij}(\tau)$ and total flows $\Sigma(\tau)$ for longer timescales $\tau$, we must include higher-order terms in the sum in Eq.~(\ref{eq:Cij_int}). For each term in the sum, we perform the corresponding integral numerically; we terminate the sum when terms become negligible. Across all analyses, we use $N = 462$ synthetic place cells to match the number identified in experiment.


\subsection{Parameter estimation.} 

Our model only contains three parameters (the average velocity $v$, the diffusion coefficient $D$, and the place field width $\sigma$), which we can estimate from the experimental data. We estimate the mean velocity $v$ and diffusion coefficient $D$ from the trajectory of the animal's position. First, we remove sections of the trajectory corresponding to the reward zone on the track, where the mouse has an abnormally low velocity (Supplementary Information). Then, for a given time interval $\tau$, we obtain the distribution of the distances that the mouse has traveled. From this distribution, we compute the mean $\hat{v}(\tau)\tau$ and variance $2\hat{D}(\tau)\tau$ in the distances traveled. We estimate the mean velocity and diffusion coefficient by averaging $\hat{v}(\tau)$ and $\hat{D}(\tau)$ over short timescales from $\tau = 0\,\text{s}$ to $\tau = 4\,\text{s}$. We arrive at the estimated values $\hat{v} = 10.2\,\text{cm/s}$ and $\hat{D} = 58\,\text{cm}^2$/s.

We estimate the place field width $\sigma$ using our place cell detection algorithm. For each place cell, the algorithm detects one or more place fields, each spanning a section of the track. The average length among all detected place fields is $42\,\text{cm}$. In our model, we approximate place fields as Gaussian with standard deviation $\sigma$, thus spanning a length of roughly $6\sigma$ along the track. We therefore estimate an average place field width of $\sigma \approx 7\,\text{cm}$.



\end{methods}

\section*{Data Availability}

The data analyzed in this paper are openly available at github.com/Kaiyue-Shi/Neural-Flow-in-the-Hippocampus.

\section*{Code Availability}

The code used to perform the analyses in this paper is openly available at github.com/Kaiyue-Shi/Neural-Flow-in-the-Hippocampus.

\begin{addendum}

\item[Supplementary Information.] Supplementary text and figures accompany this paper.

\item[Acknowledgements.] We thank J.L.~Gauthier for guiding us through the data. We thank M.P.~Leighton for helpful discussions about irreversibility. This work was supported in part by the Department of Physics and the Quantitative Biology Institute at Yale
University. 
 
\item[Author Contributions.] K.S. and C.W.L. conceived the project, designed the models, and wrote the manuscript and Supplementary Information. K.S. solved the models and performed the analyses. 
 
\item[Competing Interests.] The authors declare no competing financial interests.
 
\item[Corresponding Author.] Correspondence and requests for materials should be addressed to C.W.L. \\(christopher.lynn@yale.edu).

\end{addendum}


\newpage

\noindent {\large \myfont \textbf{Supplementary Information}}
\vspace{-30pt}

\noindent\rule{\textwidth}{.5pt}

\setcounter{figure}{0}

\section{\myfont Introduction}

In this Supplementary Information, we provide extended figures and discussion to support the results presented in the main text. In Sec.~2, we illustrate multiple mechanisms that cannot account for the decay in neural flow. Specifically, we simulate place cells with different statistical features whose total flow does not decay for a mouse running at constant velocity. In Sec.~3, we demonstrate that the total neural flow is reproduced in simulations, confirming the numerical and analytical results in the main text. Finally, in Sec.~4, we illustrate how we remove the reward zone from the mouse trajectory when esimating model parameters. \\

\section{\myfont Heterogeneity in place fields does not lead to decay in neural flow}

\begin{figure}[t!]
\centering
\includegraphics[width= 0.9\textwidth]{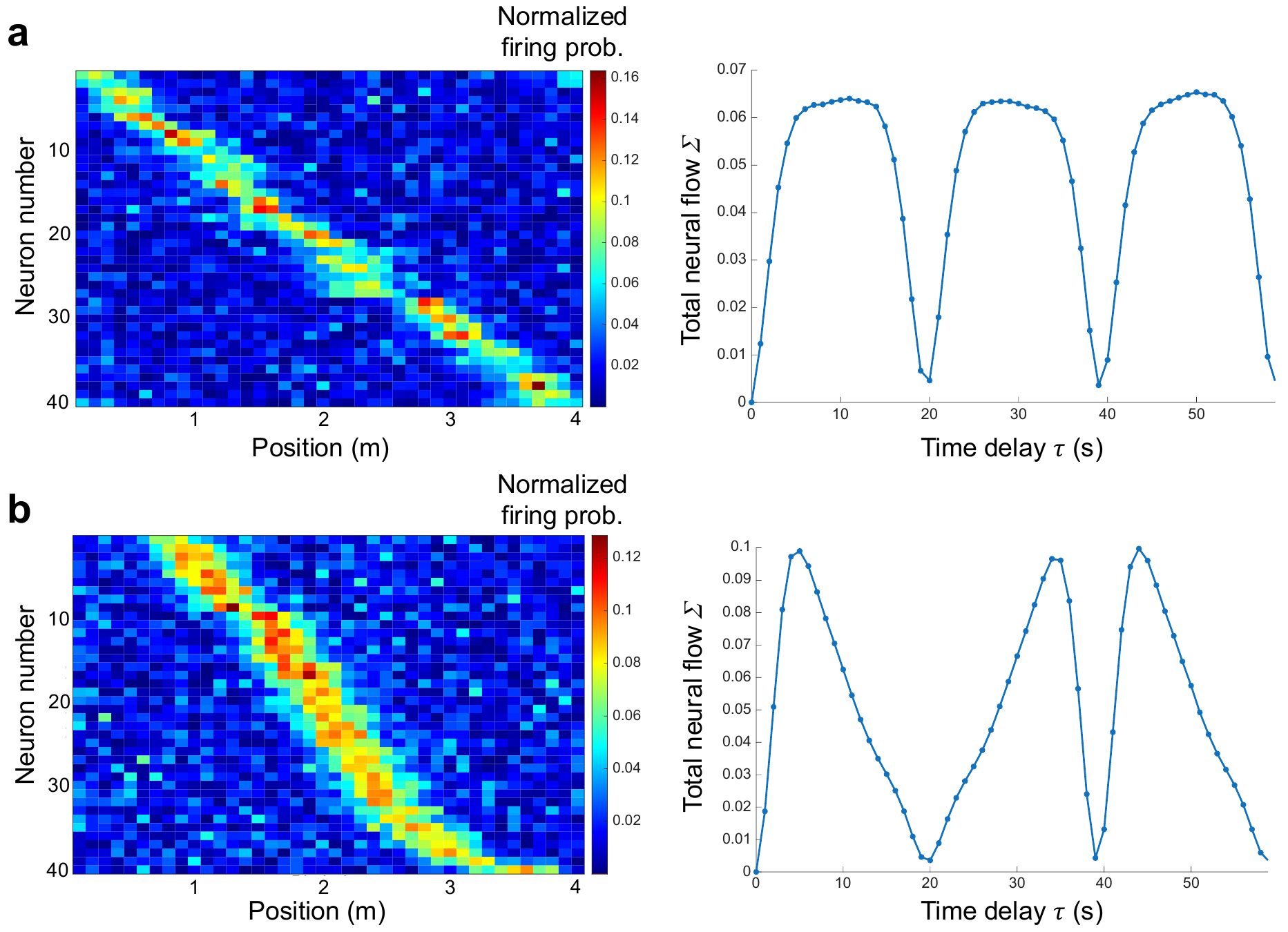}
\captionsetup{labelformat=empty}
{\spacing{1.25} \caption{\small \textbf{Fig.~S\ref{SIfig:1} $|$ Neural flow with heterogeneous place fields and constant velocity.}
\textbf{a}, Synthetic place cells are defined by noisy place fields with random centers, widths, and overall firing rates. Example conditional firing probabilities for 40 such cells (\textit{left}) and total neural flow $\Sigma(\tau)$ for $400$ cells (\textit{right}). \textbf{b}, Synthetic place cells are defined by noisy place fields with constant widths, random firing rates, and centers concentrated near the midpoint of the track. Example conditional firing probabilities for 40 such cells (\textit{left}) and total neural flow $\Sigma(\tau)$ for $400$ cells (\textit{right}). In both panels, the firing rates for each neuron are normalized such that the sum across bins equals one. Total neural flows are computed using simulations with a mouse running at constant velocity $v = 10.2\,\text{cm/s}$.
\label{SIfig:1}
}}
\end{figure}

In the main text, we show that the variation in mouse velocity leads to decay in neural flow. To rule out other possible mechanisms that could explain this decay, here we investigate models in which the mouse runs at a constant velocity (that is, with no diffusion) but with place cells that are heterogeneous.

We first study place cells with variable firing rates and noisy place fields with variable widths. To begin, as in the main text, we model the conditional firing rate of each place cell $i$ as proportional to a Gaussian $\mathcal{N}(\mu_i, \sigma_i^2)$, where the center $\mu_i$ is drawn uniformly at random along the track. Rather than setting all widths equal, for each cell $i$ we draw $\sigma_i$ uniformly at random between $0.1\,\text{m}$ and $0.3\,\text{m}$. For each place field, we then add noise $\eta=0.05 |\zeta|$ within each bin, where $\zeta \sim \mathcal{N}(0,1)$. Finally, we randomize the average firing rate of each place cell. Specifically, the average firing rate of each cell is set to match a random neuron among the top 30\% in the $1485$-neuron population. A sample of $40$ synthetic place cells is illustrated in Fig.~S\ref{SIfig:1}a. Simulating a population of $400$ such cells as the mouse runs at a constant velocity $v = 10.2\, \text{cm/s}$, we find that the total neural flow does not decay (Fig.~S\ref{SIfig:1}a, \textit{right}). In fact, the neural flow is nearly identical to the simple model with constant velocity and no heterogeneity in the place fields (Fig.~\ref{fig:3} in the main text).


We next study whether correlations in place field locations can lead to decay in neural flow. We begin by treating each place field as proportional to a Gaussian $\mathcal{N}(\mu_i, \sigma^2)$ with constant width $\sigma = 0.2\,\text{m}$. In each bin along the track, we add noise $\eta=0.05 |\zeta|$, where $\zeta \sim \mathcal{N}(0,1)$. To introduce correlations between place fields, we sample the centers $\mu_i$ from a Gaussian centered at the midpoint of the track with a standard deviation of $0.6\,\text{m}$. We again randomize the average firing rate of each cell as described above. As can be seen in Fig.~S\ref{SIfig:1}b, the place fields now cluster near the midpoint of the track. Simulating $400$ such cells as the mouse runs with constant velocity, we observe a distinct change in the total neural flow, but no decay (Fig.~S\ref{SIfig:1}b, \textit{right}). Together, these results demonstrate that the decay in neural flow is not explained by noise in the place fields, differences in firing rates, heterogeneity in place field widths, or correlations in place field locations. \\


\section{\myfont Simulations reproduce neural flow}

\begin{figure}[t]
\centering
\includegraphics[width= 0.6\textwidth]{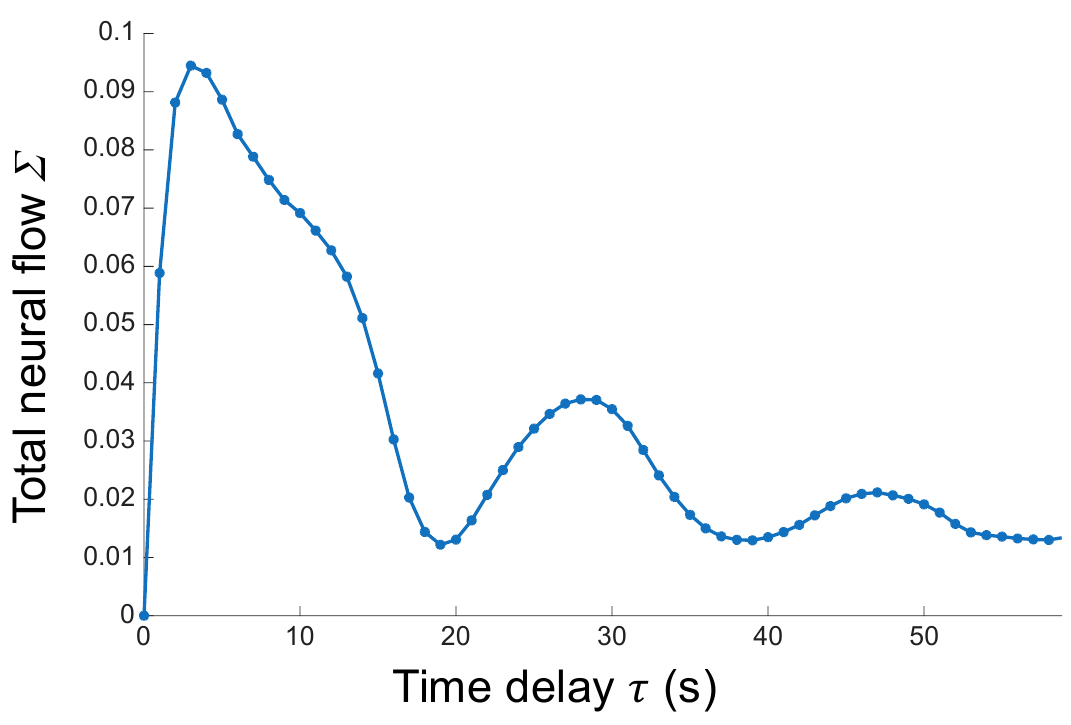}
\captionsetup{labelformat=empty}
{\spacing{1.25} \caption{\small \textbf{Fig.~S\ref{SIfig:2} $|$ Total flow in simulated neural population.} Total neural flow $\Sigma(\tau)$ computed by simulating
$462$ place cells and $1023$ non-place cells. Mouse trajectory is generated by a biased random walk. Simulations use experimental parameter values $v=10.2\,$cm/s, $D=58\,\text{cm}^2$/s, and $\sigma=7\,$cm.
\label{SIfig:2}
}}
\end{figure}

In the main text, we show that given the parameters $v$, $D$, and $\sigma$ estimated in the experiment, our model reproduces the neural flows measured in experiment. These results are all based on numerical and analytical solutions to our model. Here, we demonstrate that simulations reproduce the same results. We simulate a population of 1485 cells consisting of 462 place cells and 1023 non-place cells (thus matching the numbers in experiment). Place cells are generated as described in Sec.~2 with place fields with random centers $\mu_i$ but constant width $\sigma = 7\,\text{cm}$. Non-place cells are defined by noisy fields, whose mean firing rates are sampled from the lowest $70\%$ of firing rates within the recorded population. The trajectory of the mouse is simulated as a biased random walk,
\begin{equation}
    \label{eq:traj}
    s(t + dt) = s(t) + v\cdot dt + \sqrt{2D~ dt}\cdot\xi,
\end{equation}
where $v = 10.2\,\text{cm/s}$ and $D = 58\,\text{cm}^2$/s are set to their experimental values and $\xi \sim \mathcal{N}(0,1)$. Simulating mouse behavior and neural activity, we find that the total neural flow (Fig.~S\ref{SIfig:2}) closely matches the experimental flows described in the main text. \\

\section{\myfont Removing reward zone from mouse trajectory}

\begin{figure}[t]
\centering
\includegraphics[width=\textwidth]{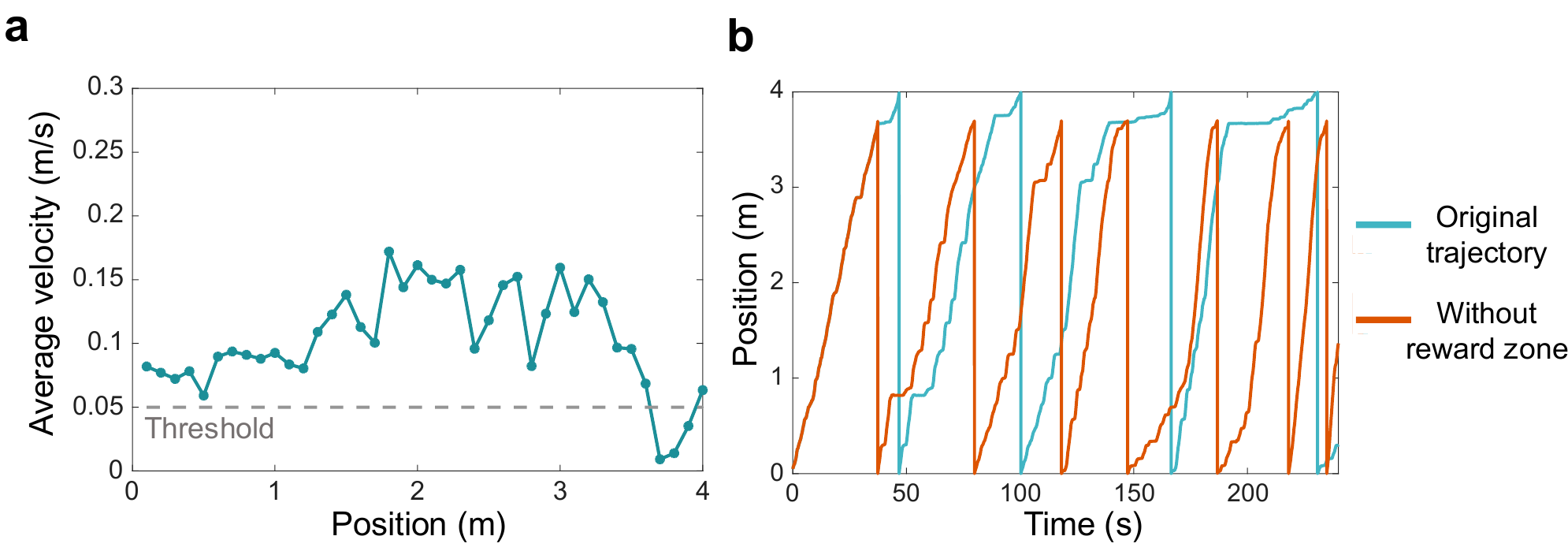}
\captionsetup{labelformat=empty}
{\spacing{1.25} \caption{\small \textbf{Fig.~S\ref{SIfig:3} $|$ Removing the reward zone from velocity and diffusion estimates.} \textbf{a}, Average mouse velocity within $40$ $0.1\,\text{m}$-wide bins along the track. The reward zone is defined between $3.6\,\text{m}$ and $3.9\,\text{m}$, where the average speed drops below $0.05\,\text{m/s}$. \textbf{b}, Mouse trajectory before (blue) and after (red) removing the reward zone. The plateaus around $3.7\,\text{m}$ in the original recording are removed, resulting in a smooth trajectory.
\label{SIfig:3}
}}
\end{figure}

In the experiment, the mouse receives a reward near the end of the track. Naturally, this section of the track (which we refer to as the reward zone) is associated with abnormally low running speeds (Fig.~S\ref{SIfig:3}a). When estimating the parameters $v$ and $D$ from the mouse behavior (as described in the main text), we remove sections of the trajectory when the mouse is in the reward zone. Specifically, we divide the track into $40$ equally-spaced bins and estimate the average velocity of the mouse within each bin (Fig.~S\ref{SIfig:3}a). The three consecutive bins with average velocities below $0.05\,\text{m/s}$ comprise the reward zone. Removing this section of the track, we arrive at a smooth trajectory from which we are able to reliably estimate the velocity $v$ and diffusion coefficient $D$ (Fig.~S\ref{SIfig:3}b).


\newpage
\clearpage

\section*{\large \myfont References}
\vspace{-30pt}
\noindent\rule{\textwidth}{.5pt}

\bibliographystyle{naturemag}
\bibliography{DetailedBalanceBib}

\newpage

\end{document}